\documentclass[]{emulateapj}

\begin{document}

\title{The Destruction of Thin Stellar Disks Via Cosmologically Common Satellite Accretion Events}

\author{
Chris W. Purcell\altaffilmark{1},
Stelios Kazantzidis\altaffilmark{2}, and
James S. Bullock\altaffilmark{1}
}

\altaffiltext{1}{
Center for Cosmology, Department of Physics and Astronomy, The University of California, Irvine, CA 92697 USA
}
\altaffiltext{2}{
Center for Cosmology and Astro-Particle Physics; and Department of Physics; and Department of Astronomy, 
The Ohio State University, Columbus, OH 43210 USA
}

\begin{abstract}  
Most Galaxy-sized systems ($M_{\rm host} \simeq 10^{12} M_{\odot}$) in the 
$\Lambda$CDM cosmology are expected to have interacted with at least one satellite with a total mass
$M_{\rm sat} \simeq 10^{11} M_{\odot} \simeq 3 M_{\rm disk}$ in the past $8$ Gyr.  
Analytic and numerical investigations suggest that this is the most precarious type of 
accretion for the survival of thin galactic disks because more massive accretion events are 
relatively rare and less massive ones preserve thin disk components.  We use high-resolution, 
dissipationless $N$-body simulations to study the response of an initially-thin, fully-formed
Milky-Way type stellar disk to these cosmologically common satellite accretion events and show that the 
thin disk does not survive.  Regardless of orbital configuration, the impacts transform the disks 
into structures that are roughly three times as thick and more than twice as kinematically 
hot as the observed dominant thin disk component of the Milky Way.  We conclude that 
if the Galactic thin disk is a representative case, then the presence of a stabilizing gas 
component is the only recourse for explaining the preponderance of disk galaxies in a 
$\Lambda$CDM~ universe; otherwise, the disk of the Milky Way must be uncommonly cold and 
thin for its luminosity, perhaps as a consequence of an unusually quiescent accretion history.

\end{abstract}

\keywords{Cosmology: theory --- galaxies: formation --- galaxies: evolution}

\maketitle

%----------------------------------
\section{Introduction} 
%----------------------------------

A solid majority of observed galaxies have disk-dominant morphology; despite wide variance in methods of sampling 
and classification, roughly 70\% of Galaxy-sized dark matter halos in the universe host late-type systems 
\citep{Weinmann_etal06,vdB_etal07,Ilbert_etal06,Choi_etal07,Park_etal07}.  Moreover, \citet{Kautsch_etal06} 
find that about one-third of all local disk galaxies have no observable pressure-supported component 
(whether a ``classical" bulge formed by the central starburst associated with a merger event, or 
a ``pseudobulge" having arisen from the secular transport 
of angular momentum towards the galactic center), and another one-third host systems with only 
pseudobulges, a conclusion supported by spheroid-disk 
decomposition of large galaxy samples \citep{Allen_etal06,Barazza_etal08}.  The vast majority
of disk stars in the Milky Way reside in the thin disk component, with an exponential
scale height of $z_d \simeq 300 \pm 60$~pc \citep[][and references therein]{Juric_etal08} and a total 
velocity dispersion of $\sigma_{\rm tot} \simeq 35$ km s$^{-1}$ \citep{Nordstrom_etal04}.  
Whether the scale height of the Galactic disk is typical for galaxies of its size is a topic of 
vital interest. Unfortunately, firm measurements for a statistical sample of galaxies have
been limited by dust obscuration, which present a problem even in K-band imaging \citep{Kregel_etal05,Yoachim_Dalcanton06}.

Aside from the considerable challenges associated with forming disk galaxies in $\Lambda$CDM 
cosmologies \citep[e.g.,][]{Mayer_etal08}, hierarchical models must also self-consistently 
maintain thin, rotationally-supported systems against the constant barrage of merging subhalos.  Though the former 
endeavor has enjoyed some recent advances \citep{Abadi_etal03,Sommer-Larsen_etal03,Brook_etal04,Robertson_etal04,
Governato_etal07}, the survival of disk galaxies during the often-violent mass accretion 
history of their dark host remains a concern \citep{TothOstriker_92,Quinn_etal93,Walker_etal96,Wyse2001} 
and has been the target of numerous studies aimed at quantifying the resilience of galactic 
disks to satellite accretion events \citep{Quinn_Goodman86,TothOstriker_92,Quinn_etal93,Walker_etal96,
Huang_Carlberg97,Sellwood_etal98,Velazquez_White99,Ardi_etal03,Hayashi_Chiba06,Kazantzidis_etal08,Read_etal08,
Villalobos_Helmi08,Hopkins_etal08}.  

Both numerical simulations \citep{Stewart_etal08} and purely analytic calculations \citep{Purcell_etal07,Zentner07} indicate 
that mass delivery into dark matter halos of mass $M_{\rm host}$ is dominated by the accretion of objects with mass $\sim (0.05-0.15) M_{\rm host}$.  
\citet{Stewart_etal08} find that $\sim 70\%$ of $10^{12} M_{\odot}$ Galaxy-sized halos 
have accreted a system of mass $M_{\rm sat} \simeq 10^{11} M_{\odot} \simeq 3 M_{\rm disk}$ into their virial radii
in the last 10 Gyr, with associated disk impacts within the last 8 Gyr.  \citet{Stewart_etal08} also find that less massive accretions are virtually 
ubiquitous, and that the merger fraction falls off quickly for satellites larger than $M_{\rm sat} \simeq 2 \times 10^{11} M_{\odot}$.
Overall, these results suggest that $\sim 1:10$ satellite accretion events represent the primary concern for disk survival in $\Lambda$CDM.  
Along these lines, recent analytic work by \citet{Hopkins_etal08} suggests that orbital energy deposition via merger 
is less destructive to a disk than was often previously surmised \citep{TothOstriker_92, Quinn_etal93,Walker_etal96}, claiming that the 
Galaxy could have undergone $\sim5-10$ independent mergers of this kind since $z\sim2$ while maintaining a thin disk.

%########################################

\begin{figure*}[!ht]
\begin{center}
\includegraphics[scale=0.7]{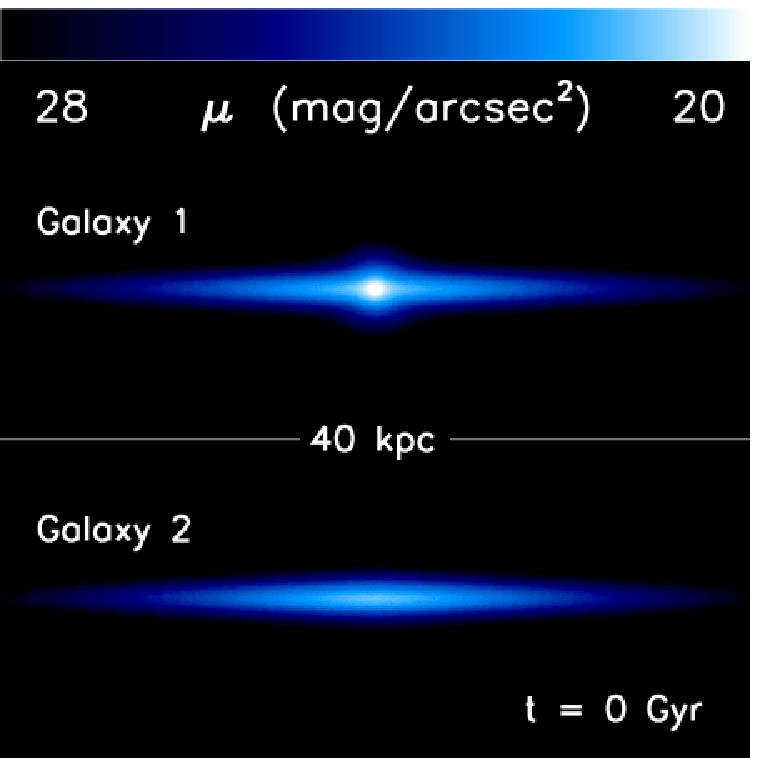}
\includegraphics[scale=0.7]{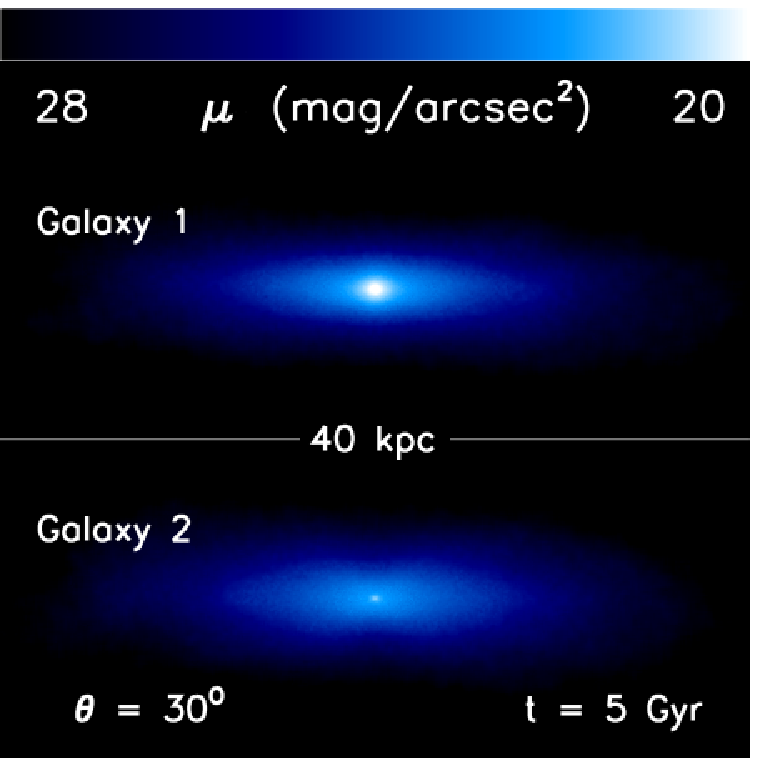} 
\includegraphics[scale=0.7]{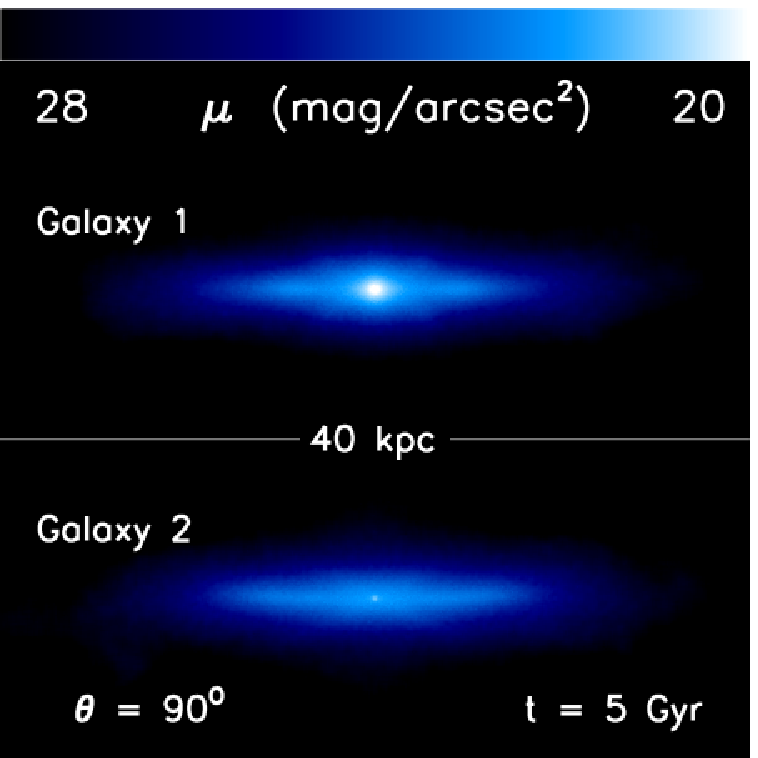} 
\end{center}    
\caption{Edge-on surface brightness maps, assuming $M_{\star}/L = 3$, for primary galaxies 1 (upper panels) and 2 
(lower panels).  Initial models ($t=0$ Gyr) are shown in the left panel, while the results 
($t=5$ Gyr) for satellite-infall orbital inclinations of $\theta = 30^{\circ}~\mathrm{and}~90^{\circ}$ appear 
in the center and right panels, respectively.
}
\label{fig:maps}
\end{figure*}
%#########################################

Recently, \citet{Kazantzidis_etal08} utilized dissipationless $N$-body simulations
to investigate the response of thin galactic disks subject to a $\Lambda$CDM-motivated 
satellite accretion history. These authors showed that the thin disk component survives, 
though it is strongly perturbed by the violent gravitational encounters with substructure. 
However, \citet{Kazantzidis_etal08} focused on infalling systems with masses 
in the range $0.2 M_{\rm disk} \lesssim M_{\rm sat} \lesssim M_{\rm disk}$, ignoring the most massive accretion events 
expected over a galaxy's lifetime. In this Letter, we expand upon this initiative by investigating the morphological and 
dynamical evolution of initially-thin Galaxy-type disks during cosmologically common $\sim 1:10$ accretion events 
involving two-component (stars and dark matter) satellites of mass $M_{\rm sat} \simeq 10^{11} M_{\odot} \simeq 3 M_{\rm disk}$. 

Working in a similar mass regime, \citet{Villalobos_Helmi08} simulated the formation of thick disks 
via the infall of satellite galaxies with virial masses $\sim 10-20\%$ that of the host, using both a $z=0$ Galactic 
primary system and a scaled version at $z=1$ in order to show that realistic thick disks result from these impacts. 
Though our preparation is similar, our goals and techniques are different.  We aim to determine whether {\em any} thin, 
dynamically cold component can survive such an event, and conservatively use a primary disk that is as massive as the Milky Way disk 
{\em today}.  

Past investigations into the stability of galactic disks against the infall of satellites have often 
suffered from the necessities of numerical limitations or from analytic axioms later deemed 
incompatible with standard cosmological models; for example, the modeling of one or more structural components as 
rigid potentials \citep{Quinn_Goodman86,Quinn_etal93,Sellwood_etal98,Ardi_etal03,Hayashi_Chiba06}, the initialization of 
a disk much thicker than the old, thin stellar disk of the Galaxy \citep{Quinn_etal93,Walker_etal96,Huang_Carlberg97,
Velazquez_White99,Font_etal01,Villalobos_Helmi08}, the infall of satellites with only a concentrated baryonic component 
\citep{Quinn_etal93,Walker_etal96,Huang_Carlberg97,Velazquez_White99}, and the imposition of subhalo infalls with 
orbital parameters inconsistent with $\Lambda$CDM cosmological models \citep{Quinn_etal93,Walker_etal96,Huang_Carlberg97}.  
Analytic arguments, meanwhile, have historically been forced to assume simplifications such as the local deposition 
of a satellite's orbital energy \citep{TothOstriker_92}, or the absence of global heating modes \citep{Benson_etal04,Hopkins_etal08} which 
are analytically shown to dominate disk heating by \citet{Sellwood_etal98}, although the latter authors employ a rigid satellite model and 
perfectly radial polar orbits for their simulated experimental tests.  Fortunately, advances both in computational power and 
in our understanding of $\Lambda$CDM expectations allow us to address these concerns directly.

Our contribution improves upon earlier studies in several important respects.  First and foremost, we examine the 
response of galactic disks to accretion events that represent the primary concern for disk survival in $\Lambda$CDM 
cosmologies. Secondly, we employ galaxy and satellite models that are constructed in equilibrium from fully 
self-consistent distribution functions and which have the resolution in force and mass to study the heating of a disk that is as 
thin as the old thin stellar disk of the Milky Way ($z_d \simeq 300$~pc); in synergy with the high mass and force resolution we adopt, this quality 
allows us to construct equilibrium $N$-body models of disk galaxies that are as thin as the {\it old}, thin stellar 
disk of the Galaxy.  Lastly, the masses, density structure, stellar content, and orbital configurations of our infalling satellites 
are directly motivated by the prevailing $\Lambda$CDM paradigm of structure formation.

%----------------------------------------
\section{Methods}
\label{sec:methods}
%----------------------------------------

All simulations are performed using the multi-stepping, parallel, 
tree $N$-body code PKDGRAV \citet{Stadel2001}, in which we set the gravitational softening 
length to $\epsilon =$ 100 pc and 50 pc for dark matter and stellar particles, respectively.

%----------------------------------------
\subsection{Primary and Satellite Galaxy Models}
\label{subsec:models}
%----------------------------------------

We construct $N$-body realizations of primary disk galaxies and satellites using the 
method of \citet{Widrow_etal08}. This technique produces self-consistent, multi-component 
galaxy models that are ideal for studying complex dynamical processes associated with the 
intrinsic fragility of galactic disks such as gravitational interactions with infalling 
subhalos.  We explore two initial models for the primary galaxy in our satellite-disk encounter 
simulations: {\em Galaxy 1} (hereafter G1), a Milky-Way-analog system drawn from the set of 
self-consistent equilibrium models that best fit Galactic observational parameters as produced 
by \citet{Widrow_etal08}; and {\em Galaxy 2} (hereafter G2), an identical system save for the absence of a central bulge, i.e., the two models 
have stellar disks and dark halos with equivalent {\it initial} properties.  
In each case the dark matter halo of the primary galaxy was populated by $4 \times 10^6$ particles following the \citet[][hereafter NFW]{Navarro_etal96} 
density profile with scale radius $r_s = 14.4$ kpc, and the bulge in G1 (comprised of $5 \times 10^5$ particles) 
contained a stellar mass $M_{\rm bulge} = 9.5 \times 10^9 M_{\odot}$ following a S\'ersic profile with effective radius $R_e = 0.58$ kpc and index $n = 1.118$. The stellar disks, comprised of $10^6$ particles each, contained a mass $M_{\rm disk} = 3.6 \times 10^{10} M_{\odot}$ following an exponential distribution in cylindrical radius 
with scale length $R_d = 2.84$ kpc, while the vertical distribution of stars was described by a sech$^2$ function with 
$z_d = 0.43$ kpc being the vertical scale height.  We note that the choice of numerical and physical parameters 
minimize secular evolution (e.g., strong bar formation, artificial heating through interactions with massive halo particles) on the 
timescales of relevance to our investigation, which could interfere with the interpretation of our results. In the left panel of 
Figure~\ref{fig:maps}, we show the edge-on surface brightness map for both primary galaxy models, having assumed 
a stellar mass-to-light ratio $M_{\star}/L = 3$. The satellite galaxy in each case was initialized with $9 \times 10^5$ dark particles 
representing a mass $M_{\rm sat} = 1.0 \times 10^{11} M_{\odot}$ within the virial radius of a halo which is well-fit by an 
NFW profile with a concentration $c_{\rm vir} \simeq 14$ at $z=0.5$.  We populate this satellite with a stellar mass 
$M_{\star} = 2.2 \times 10^9 M_{\odot}$, roughly corresponding to the upper-$1\sigma$ limit derived by 
\citet{Conroy_Wechsler08} for $M_{\star}/M_{\rm sat}(z \sim 0.5)$ at our subhalo's virial mass, and we distribute these 
$10^5$ stellar particles in a central spheroid with S\'ersic index $n \sim 0.5$ according to the distribution of shape parameters 
versus dwarf elliptical galaxy magnitudes found by \citet{vanZee_etal04} in their survey of Virgo cluster members.

%----------------------------------------
\subsection{Satellite Galaxy Orbits}
\label{subsec:orbits}
 %----------------------------------------

Our initial subhalo velocity vectors are motivated by cosmological investigations of substructure accretions, where
the distributions of radial and tangential velocity components ($v_r$ and $v_t$) peak 
respectively at 90\% and 60\% of the virial velocity of the satellite's host halo \citep{Benson2005,Khochfar_Burkert06}.
In our case this corresponds to an initial subhalo velocity with $v_r = 116$ km/s and $v_t = 77$ km/s.  
We initiate the infall of each simulation's subhalo at a relatively large radius of approximately $120$ kpc to 
ensure that the disk does not suffer substantial perturbations due to the sudden presence of the satellite's potential well.  
We simulate an array of orbital inclinations ($\theta = 0^{\circ}, 30^{\circ}, 60^{\circ}, \mathrm{and}~90^{\circ}$, defining 
$\theta$ as the angle between the angular momentum axes of the disk and the orbit) in order to assess the consequence 
of this parameter on the evolution of the galactic disk.  In the polar infall ($\theta = 90^{\circ}$), we eliminate the tangential 
velocity component of the subhalo, sending the satellite on a direct-impact trajectory into the center of the primary galaxy; 
this case is somewhat unrealistic, but provides an interesting experimental benchmark.
All but one of the non-polar subhalo orbits are initialized as prograde with respect to the primary galaxy's rotation; 
we also simulate a retrograde orbit for G1 with $\theta = 60^\circ$ in order to investigate whether the heating 
effects are reduced \citep[as conjectured by][]{Velazquez_White99}. All simulations were evolved for a total of 5 Gyr, 
after which the subhalo has fully coalesced into the center of the host halo and the stellar disk has 
relaxed into stability; although there are certainly remnant features in the outer disk and halo that 
will continue to phase-mix and virialize on a much longer timescale, our investigations indicate 
that the disk-evolution process has reached a quasi-steady state by this point in the encounter's evolution.

%%########################################
\begin{figure}[t]
\centerline{\epsfxsize=3.5in \epsffile{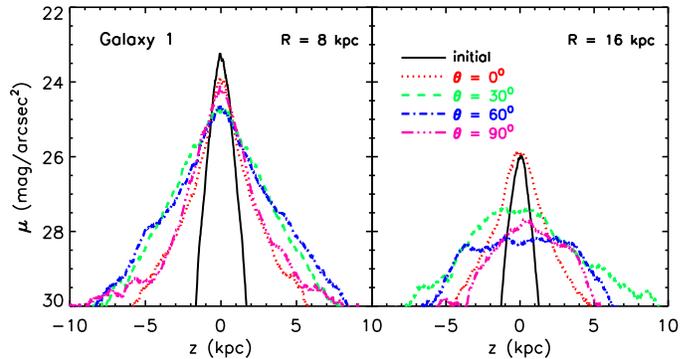}}
\caption{Minor-axis surface brightness profiles for initial and final models at two Galactocentric 
radii: $R=R_{\odot}=8~\mathrm{kpc}$ (left panel) and $R=2R_{\odot}=16~\mathrm{kpc}$ (right panel).
}
\label{fig:diskheight}
\end{figure}
%%#########################################

%%########################################
\begin{figure}[t]
\centerline{\epsfxsize=3.2in \epsffile{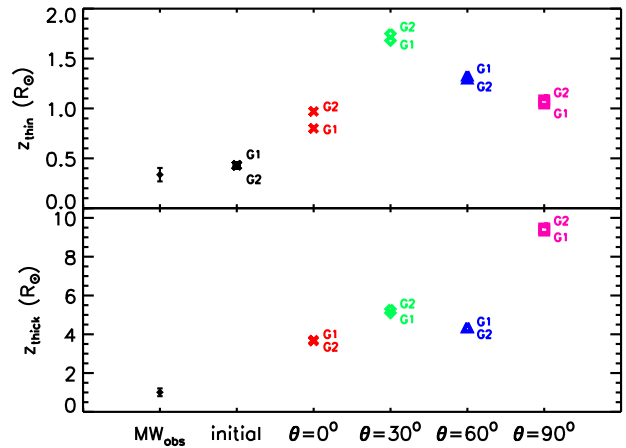}}
\caption{The thin- and thick-disk scale heights in the final state ($t=5$ Gyr) for each of our simulated 
galaxies, compared to the values derived by \citet{Juric_etal08} for the Milky Way. The two panels 
show the result of a two-component sech$^2$ fit, with the upper (lower) panel describing the thin (thick) 
disk's scale height.}
\label{fig:scaleheight}
\end{figure}
%%#########################################

\begin{table*}[t] 
\centering
\begin{minipage}{\textwidth}
\centering
\tabcolsep 11pt
  \caption{Final ($t=5$ Gyr) Galaxy Properties at $R=R_{\odot}=8$ kpc}
 \label{finalscaleheights}
 \begin{tabular}{@{}lccccccc}
  \hline
  \hline
  Orbital       & $z_{\rm thin}$            & $z_{\rm thick}$   & $<|z|>$       & $z_{\rm median}$  & $R_d$         & $\sigma_{z}$  & $\sigma_{\rm tot}$ \\
  Inclination           & (G1, G2;              &(G1, G2;       &(G1, G2;           &(G1, G2;            &(G1, G2;       &(G1, G2;                &(G1, G2;\\
  of Subhalo                & in kpc)                 & in kpc)         & in kpc           & in kpc)            & in kpc)        & in km/s)            & in km/s) \\
  \hline         
   initial ($t=0$ Gyr)                                   & $0.43, 0.43$               & N/A                       & $0.3, 0.3$    & $0.3, 0.3$      & $3.0, 3.0$             & $19.1, 18.7$     & $50.9, 52.1$\\
  $\theta = 0^{\circ}$   (prograde)             & $0.80, 0.97$               & $3.70, 3.65$               & $0.9, 1.0$    & $0.5, 0.6$          & $2.3, 4.5$              & $25.1, 28.0$        & $115.2, 107.1$\\
  $\theta = 30^{\circ}$ (prograde)             & $1.68, 1.75$               & $5.10, 5.30$                & $1.7, 1.8$    & $1.0, 1.1$          & $3.5, 2.9$              & $37.9, 40.6$        & $95.5, 102.6$\\
  $\theta = 60^{\circ}$ (prograde)             & $1.33, 1.30$               & $4.38, 4.35$                & $1.8, 2.0$    & $0.9, 1.1$          & $2.2, 2.6$              & $33.5, 35.1$        & $82.4, 86.1$\\
  $\theta = 60^{\circ}$-retro (G1 only)      & $1.18$                        & $6.50$                         & $2.1$           & $0.8$             & $2.6$                    & $31.4$              & $83.3$\\
  $\theta = 90^{\circ}$ (polar)                   & $1.05, 1.08$               & $9.35, 9.45$             & $2.0, 1.9$    & $0.6, 0.7$          & $4.2, 3.0$              & $26.2, 29.4$        & $70.0, 75.6$\\
  \hline
  Milky Way (observed)\footnote{For the Galaxy's empirical constraints, we quote the disk scale heights and lengths derived by \citet{Juric_etal08} and the velocity dispersions obtained by 
\citet{Nordstrom_etal04} for solar-neighborhood stars of median age ($t \sim 2-3$ Gyr).}                & $0.34$                            & $1.01$            & $0.298$           & $0.208$             &$2.6, 3.6$           & $\sim 10-20$             & $\sim 30-40$\\
&  &  &  &  &(thin, thick) & & \\
   \hline
  \hline
 \end{tabular}
 \end{minipage}
\end{table*}

%-------------------------------------------------
 \section{Results and Implications}
\label{sec:results}
%-------------------------------------------------

Edge-on surface brightness profiles for remnants of the $\theta = 30^{\circ}$ and 90$^\circ$ impacts are shown in the middle and left
panels of Figure 1, where the upper and lower renderings correspond to primary cases G1 and G2, respectively
(with and without initial bulge).   It is clear from these images that the resultant disks are considerably thicker than the
initial case. We note that while the stars in the accreted satellite end up in the final disk remnant 
(c.f.~Villalobos \& Helmi 2008), primary disk stars dominate these images, even high above the plane. 

Figure~\ref{fig:diskheight} shows the minor-axis surface brightness profiles for the G1 simulations using $M_{\star}/L = 3$.  The left 
panel shows a vertical slice at a projected radius of $R_{\odot}=8$ kpc and the right panel shows a similar slice at
radius $2 \, R_{\odot}$.  Black solid lines show the initial disk and different color/line types represent the remnants as indicated.  
Clearly, the resultant disks are dramatically thicker than the initial galaxy model in each case.  In order to conservatively
compare our disks to the Milky Way, we allow for thick {\em and} thin components by fitting 
a double-sech$^2$ profile at $R_{\odot}$.  The fitted scale heights are shown in Table~\ref{finalscaleheights}
and compared directly in Figure~\ref{fig:scaleheight} to the Galactic values obtained by \citet{Juric_etal08}
\footnote{The scale heights derived by \citet{Juric_etal08} belonged 
to exponential profiles; we have therefore multiplied these values by a factor of 1.12 
to obtain scale heights belonging to sech$^2$ profiles that fall by one mag/arcsec$^2$ at the same height as the 
exponential fits.  This multiplicative factor is more appropriate for thin-disk comparisons
near the peak of the profile than the widely-used factor
of 2 that matches exponential and sech$^2$ profiles at large
heights above the disk plane.}.  Though we initially employ a disk that is thicker ($z_{\rm thin} = 0.43$ kpc) and therefore
conservatively more robust to accretion events (Kazantzidis et al. 2009, in preparation) than the Galactic value of 
$z_{\rm thin} = 0.34$ kpc from \citet{Juric_etal08}, the final systems all have thin-disk components with $z_{\rm thin}$ 
larger by a factor of $\sim 3-5$ than the Milky Way. Moreover, the low-surface-brightness thick component in our remnant
disks is also considerably thicker than the Galactic thick disk, with scale heights so large ($z_{\rm thick} 
\simeq 4 -10$ kpc) that this material would likely be considered a stellar halo component.

\begin{figure}[b]
\begin{center}
\centerline{\epsfxsize=3.5in \epsffile{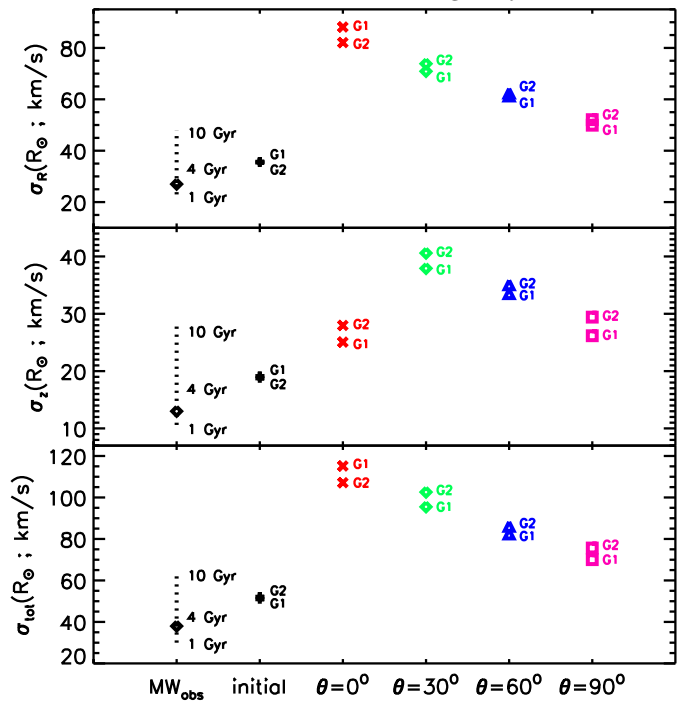}}
\end{center}
\caption{The radial and vertical components of velocity dispersion $\sigma_R$ and $\sigma_z$ (top and middle panels), 
as well as the total stellar dynamical temperature $\sigma_{\rm tot}$, at the solar neighborhood ($R_{\odot}=8$ kpc) 
of our simulated disks, compared to the local values obtained by the Geneva-Copenhagen survey results described 
in \citet{Nordstrom_etal04}.  In each coordinate, the observational spread is marked by a {\em dotted line} and 
the dispersion of the sample's median-age stars ($t \sim 2-3$ Gyr) is denoted by a {\em diamond}.  }
\label{fig:sigma}
\end{figure}

A second relevant measure of disk survival is the stellar velocity dispersion; we therefore compare the velocity ellipsoid 
of our final disks to that observed in the solar neighborhood by the Geneva-Copenhagen Survey 
\citep[][see also \citealt{Seabroke_Gilmore07}]{Nordstrom_etal04}.
In Figure~\ref{fig:sigma}, we show these values for velocity dispersion in each coordinate, where the
indicated range spans the stellar population age, and the point is placed at the median age 
$t \sim 2-3$ Gyr according to \citet{Nordstrom_etal04} for this local sample.  Shown also are 
the corresponding dispersion components for our initial and final stellar disks measured within an 0.3-kpc box centered 
on the disk plane at $R_{\odot} = 8$kpc.  As summarized in Table 2, each of our simulated merger remnants are 
substantially enhanced in all three components of velocity dispersion ($\sigma_{R,\phi,z}$ corresponding to 
$\sigma_{U,V,W}$).  The total dispersion $\sigma_{\rm tot}=(\sigma_R^2+\sigma_{\phi}^2+\sigma_z^2)^{1/2}$ 
increases by a factor of $\sim 1.5-2$ compared to that of the initial disk.

%%########################################
%%#########################################

%-------------------------------------------------
\section{Conclusions and Discussion}
\label{sec:discuss}
%-------------------------------------------------

Using fully self-consistent $N$-body simulations of satellite-disk interactions we have quantitatively demonstrated for the first time that 
cosmologically common accretion events of 
mass ratio $\sim 1:10$ do not preserve thin, dynamically cold stellar disks like the old, thin stellar disk of the Milky Way.  
This has potentially serious ramifications for models of galaxy formation and evolution. 
It is possible that our benchmark case of the Milky Way is not representative, and that the
Galaxy sits within a rare halo that has not experienced in the last $\sim 8$ Gyr a disk impact 
associated with a significant accretion event, as posited in the observationally-motivated suggestion of \citet{Hammer_etal07}, 
in which the Galaxy is shown to have remarkably low angular momentum and stellar mass compared to 
local spiral galaxies in host halos of similar mass.  Future investigations may help quantify the range of thin-disk 
scale heights in the local universe.

Otherwise, the addition of gas physics may play a role in explaining the apparent discrepancy.
Gas can cool and reform a thin disk, and its presence may stabilize the stellar disk \citep[e.g.,~][]{Robertson_etal06}.  
The regrowth of the massive thin disk {\em after} a satellite accretion may cause heated stars to contract and lose kinetic energy.  
Accurate treatment of the various aspects of hydrodynamics will therefore play a crucial role in the capacity of simulated galaxy 
evolution to reproduce thin disks such as those that dominate observed galaxy catalogs.

In a recent paper \citet{Hopkins_etal08} have argued that disk heating is less effective than previously
thought and that the expected merger histories of $\Lambda$CDM~halos are compatible with the high thin-disk fraction
seen in the Universe.  It is important, therefore, to investigate this point of disagreement.  Their result, a reshaping of the 
arguments presented in \citet{TothOstriker_92} (updated to reflect the more realistically radialized orbits of a $\Lambda$CDM 
cosmology), relied primarily on an analytic formula, normalized to simulations with much lower mass and force resolution than those 
explored here, to map the ratio ($M_{\rm sat}/M_{\rm host}$)  to a disk heating parameter $\Delta{}H/R$, where $H$ is the median scale 
height of the resultant disk and $R$ is the radius where the height is measured (and must be within a factor of two of the disk 
half-mass radius $R_e$).  For the $\sim 1:10$ mass-ratio accretion events we explore here, the \citet{Hopkins_etal08} formula predicts a 
disk thickening of $\Delta{}H/R \simeq 0.015$.   Our simulations typically exhibit significantly more heating; at $R=R_e$ we measure
$\Delta{}H/R \simeq (0.03 - 0.09)$ and, because of the impact-induced flaring, we measure
even larger values $\Delta{}H/R \simeq (0.05 - 0.11)$ at $R = 2R_e$.  

It is perhaps not surprising that our results disagree with first-order 
analytic expectations.  In addition to direct heating, the resultant disk structure is affected by global modes such as bending 
and density waves excited in the disk as the interaction
occurs \citep{Sellwood_etal98}, and not included in the simple analytic scalings is a dependence on the orbital inclination of the encounter
that is likely associated with resonant coupling.  Finally, though \citet{Hopkins_etal08} normalized their results to numerical simulations, 
those initial disks were significantly thicker than the Galactic-type disk we have simulated, and were therefore more robust to tidal 
perturbations.  Direct 
numerical experiments involving satellite-disk encounters indicate that mass-ratio, orbital inclination, initial disk scale height, and relative dark matter fraction 
are all crucial in determining the degree to which galactic disks are perturbed by infalling subhalos (Figures~\ref{fig:scaleheight},~\ref{fig:sigma}, 
and Kazantzidis et al. 2009, in preparation).  More detailed analysis is forthcoming of the morphological and dynamical effects experienced by 
our disks; among other concerns, we defer for future work the issues of stellar-halo/thick-disk distinguishability and the reinforcement of central bulges by accreted stars. 

\acknowledgments

We thank Joachim Stadel for providing the PKDGRAV code.
We would like to thank Charlie Conroy, Phil Hopkins, Kyle Stewart, and Andrew Zentner for useful discussions
as well as Larry Widrow and John Dubinski for kindly making available 
the software used to set up the primary galaxy models. 
CWP and JSB are supported by National Science Foundation (NSF) 
grants AST-0607377 and AST-0507816, and the Center for 
Cosmology at UC Irvine.  SK is supported by the Center for Cosmology and 
Astro-Particle Physics at The Ohio State University. The numerical simulations 
were performed on the IA-64 cluster at the San Diego Supercomputing Center.

\end{document}